\documentclass[ aip, pop, reprint, amsmath,amssymb,floatfix,hyphens,break]{revtex4-1}
\usepackage{graphicx, bm, color} 
\usepackage{siunitx}

\usepackage{dcolumn}

\usepackage{textgreek} 
\usepackage{etoolbox}

\def\twod/{2D(3$v$)}
\def\um/{\textmu m} 
\newcommand{\Wcm}[2][0]{\ifnumcomp{0}{=}{#1}{}{$#1~\times$~}$10^{#2}$~W~cm$^{-2}$} 
\def\mildI/{\Wcm[5.4]{17}}
\def\relI/{\Wcm[3]{18}}

\begin{document}

\title{Particle-in-Cell Simulations of Density Peak Formation and Ion Acceleration from Short Pulse Laser-Driven Ponderomotive Steepening}
\author{Joseph R. Smith}
 \email{smith.10838@osu.edu}
\affiliation{\mbox{Department of Physics, The Ohio State University, Columbus, OH, 43210, USA}} 

\author{Chris Orban}
 \affiliation{\mbox{Department of Physics, The Ohio State University, Columbus, OH, 43210, USA}} 
 \affiliation{Innovative Scientific Solutions, Inc., Dayton, OH, 45459, USA}
 
\author{Gregory K. Ngirmang}
\affiliation{Innovative Scientific Solutions, Inc., Dayton, OH, 45459, USA}

\author{John T. Morrison}
\affiliation{Innovative Scientific Solutions, Inc., Dayton, OH, 45459, USA}

\author{Kevin M. George}
\affiliation{Innovative Scientific Solutions, Inc., Dayton, OH, 45459, USA}

\author{Enam A. Chowdhury}
\affiliation{\mbox{Department of Physics, The Ohio State University, Columbus, OH, 43210, USA}} 
\affiliation{Intense Energy Solutions, LLC., Plain City, OH, 43064, USA}

\author{W. M. Roquemore}
\affiliation{Air Force Research Laboratory, Dayton, OH, 45433, USA}

\date{\today}

\begin{abstract}

We use particle-in-cell (PIC) simulations and simple analytic models to investigate the laser-plasma interaction known as ponderomotive steepening. When normally incident laser light reflects at the critical surface of a plasma, the resulting standing electromagnetic wave modifies the electron density profile via the ponderomotive force, which creates peaks in the electron density separated by approximately half of the laser wavelength. What is less well studied is how this charge imbalance accelerates ions towards the electron density peaks, modifying the ion density profile of the plasma. Idealized PIC simulations with an extended underdense plasma shelf are used to isolate the dynamics of ion density peak growth for a 42~fs pulse from an 800~nm laser with an intensity of 10$^{18}$ W~cm$^{-2}$. These simulations exhibit sustained longitudinal electric fields of 200~GV m$^{-1}$, which produce counter-steaming populations of ions reaching a few keV in energy. We compare these simulations to theoretical models, and we explore how ion energy depends on factors such as the plasma density and the laser wavelength, pulse duration, and intensity. We also provide relations for the strength of longitudinal electric fields and an approximate timescale for the density peaks to develop. These conclusions may be useful investigating the phenomenon of ponderomotive steepening as advances in laser technology allow shorter and more intense pulses to be produced at various wavelengths. We also discuss the parallels with other work studying the interference from two counter-propagating laser pulses.

\end{abstract}

\maketitle


\section{Introduction}
\label{sec:intro}

\begin{figure*}
\includegraphics{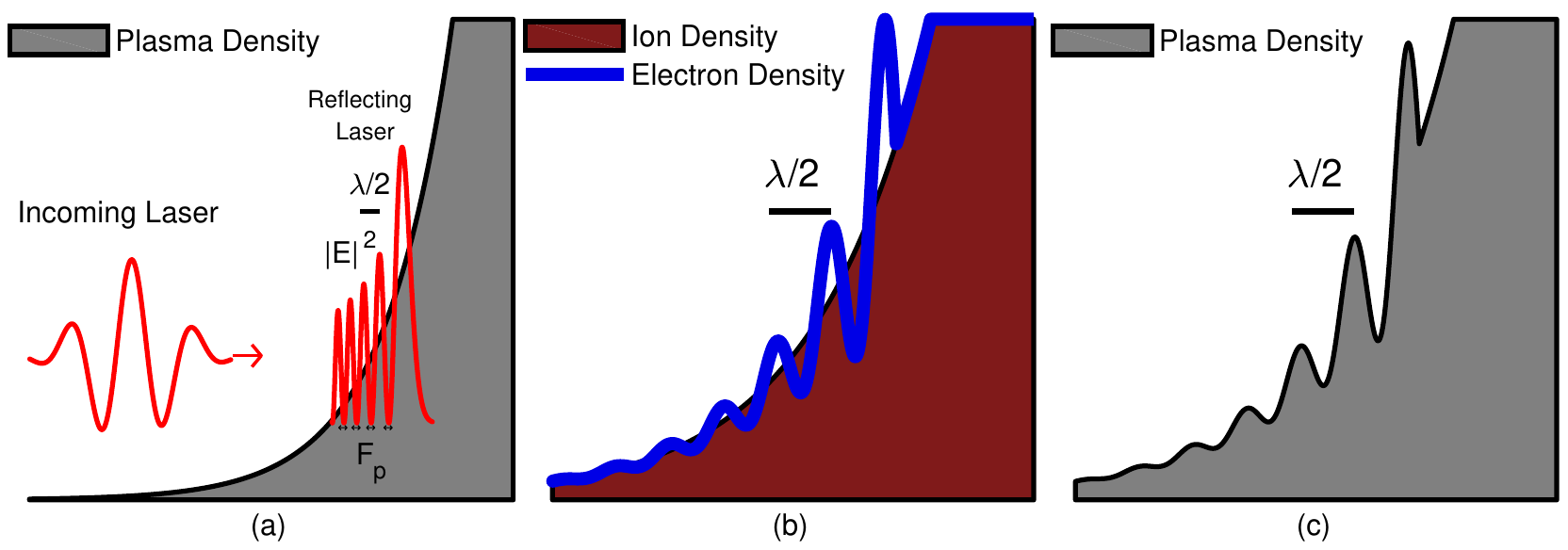}
\caption{Sketch of the ponderomotive steepening process. As illustrated in (a), a normally incident laser pulse reflects at the critical density of a plasma and forms a standing electromagnetic wave. This causes the electrons to form peaks near the extrema (separated by $\approx \lambda/2$) of the standing wave via the ponderomotive force (b). The modification of the electron density creates a charge imbalance (sustained by the standing wave), which accelerates ions towards the electron peaks. In time, this modifies the density of the plasma as illustrated in~(c). Note that (b) and (c) only include the standing wave region from (a). } \label{fig:sketch}
\end{figure*}

Ultra-intense laser interactions with dense targets represent an interesting regime, both from a fundamental and an applied perspective, that has not yet been exhaustively explored. One less explored phenomenon in this regime is the formation of electron and ion density peaks due to a laser pulse that strongly reflects from a dense target. There are papers that discuss this process  -- sometimes called ponderomotive steepening -- going back to Estabrook et al.~1975\cite{estabrook1975two}. Figure~\ref{fig:sketch} provides a qualitative sketch of the physics involved in this laser-plasma interaction. First, a normally incident, linearly polarized laser makes a strong reflection from a dense plasma. The interference between the incident and reflected pulse produces a standing wave pattern (Fig.~\ref{fig:sketch}a). The ponderomotive force associated with this standing wave has a strong effect on the electron distribution (Fig.~\ref{fig:sketch}b) and, over time, peaks form in the density of both the electrons and ions (Fig.~\ref{fig:sketch}c). Readers who are familiar with Kruer's 1988 textbook \cite{kruerbook} will recall the discussion of this phenomenon there. Ponderomotive steepening also draws many parallels to theoretical and computational work that considers the standing electromagnetic (EM) wave formed by crossing two laser pulses to generate plasma optics such as plasma gratings\cite{plaja199diffraction, Sheng2003} and so-called transient plasma photonic crystals\cite{Lehmann2016,Lehmann2017,Lehmann2019} which are phenomena that may have useful applications in the future (see discussions in Refs.~\cite{Sheng2003,Lehmann2017}). From an experimental point of view, ponderomotive steepening only requires one laser pulse, and the high densities near the critical surface allow for larger transverse electric fields than with counter-propagating lasers in low density media.

We are motivated to return to this topic with fresh eyes in part due to the maturation of technologies to produce intense laser pulses at mid-infrared (IR) wavelengths ($2$~\si{\um} $\lesssim \lambda \lesssim  10$~\si{\um}) \cite{lasermag}. This presents an opportunity to examine the wavelength dependence of intense laser-matter interactions to see if theoretical models developed from studying laser interactions at shorter wavelengths remain valid at longer wavelengths (e.g.~Ref.~\cite{ngirmang2017particle}, and ongoing research efforts\cite{MURIresearch}). As discussed later, the density peaks that form with ponderomotive steepening are separated by approximately half the laser wavelength. It is therefore challenging to detect and resolve these density peaks in near-IR or shorter-wavelength laser interactions. There have been many experiments that confirm that the ponderomotive force does steepen the plasma profile near the target as expected (e.g~\citet{Fedosejevs_etal1977,Gong_etal2016}) and researchers have found evidence in experiments with counter-propagating near-IR laser pulses that the interference shapes the plasma distribution in a low density medium (e.g.~\citet{suntsov2009femtosecond}). However, multiply-peaked ponderomotive steepening  has not yet been directly observed with interferometry or by other means.  We aim to provide useful analytic insights for experimentalists working to demonstrate this effect.

A challenge for connecting theory to observation is that ponderomotive steepening is simplest to model and has larger longitudinal electric field strengths when the laser interactions are at normal incidence, whereas at the highest intensities normal incidence experiments are rare because of the potential damage that the reflected pulse could do to optical elements. There are, however, methods to protect optics from the reflected pulse. Normal incidence experiments were conducted, for example, at $\approx 10^{18}$~W~cm$^{-2}$ peak intensities with $\approx 3$~mJ pulses at a kHz repetition rate in Refs.~\cite{Morrison_etal2015,Feister_2017}. Although the present paper is not tied to modeling interactions from a particular laser system, it is important to note that normal incidence experiments can be performed. As will be discussed, ponderomotive steepening is not typically thought of as an ion acceleration process but some ions do reach significant energies due to the charge separation caused by the ponderomotive force; and experiments could investigate this regime. According to estimates that agree with our 2D(3v) PIC simulations, under the right conditions and laser parameters these interactions have the potential to accelerate ions to energies exceeding 100~keV. Experiments of this kind would also be interesting as a new type of code validation experiment for high intensity laser-plasma interactions. Both during and after the laser interaction, ions move and the electron and ion density profiles change over time which can be investigated with interferometry\cite{Feister_etal2014,Grava_etal2008, Filevich_etal2009} and measurements of escaping ion energies (e.g.~Ref.~\cite{morrison_etal2018}). The simplicity and symmetry of normal incidence interactions would be helpful for comparing experiment to simulation and theory in a straightforward way.

In Sec.~\ref{sec:theory}, we provide a brief review of the physics of ponderomotive steepening and identify the relevant timescales for ion motion using simple analytic models. In Sec.~\ref{sec:sims} we describe 2D(3$v$) PIC simulations that exhibit multiply-peaked ponderomotive steepening. In Sec.~\ref{sec:results} the simulation results are presented and compared to the analytic models discussed in Sec.~\ref{sec:theory}. Finally, we address implications of our results in the concluding sections.

\section{Ponderomotive Steepening and Ion Acceleration}\label{sec:theory}
The traditional analytic approach for ponderomotive steepening considers a steady state solution to the fluid equations, to which a term for the ponderomotive force is added. The electric field is then assumed to take a particular form based on the geometry of the problem and to allow for numerical solutions or approximate solutions~\cite{estabrook1975two,Lee_1977, Jones_1981,Estabrook_1983,wenda1988ponderomotive,kruerbook}. These approximations limit the validity of the conclusions, and the steady state solution provides little insight into the dynamics of the phenomenon. We investigate these dynamics by developing a simple model to estimate the longitudinal electric fields experienced by the ions and comparing the predictions to PIC simulations. Our simple model is similar in many ways to an analytic model described in a recent paper by \citet{Lehmann2019} that considers the dynamics of the electron motion for the case of lower intensity counter-propagating laser beams in a low density medium with 1D Vlasov simulations. Our paper is complimentary to theirs because we consider standing waves that form from the normal incidence reflection of intense laser pulses from an overdense target preceded by a $\approx$1/20th of critical density shelf and as just mentioned, we focus on the dynamics of the ions. Our 2D(3$v$) PIC simulations also include the focusing of the laser. Where appropriate, we provide comments for those wishing to compare our work to \citet{Lehmann2019}. The timescales and intensity thresholds we develop are very similar to their models.

\subsection{A simple model for ponderomotive and electrostatic forces in ponderomotive steepening}
As sketched in Fig.~\ref{fig:sketch}, the laser creates a charge imbalance due to the ponderomotive force on the electrons, which in turn creates a longitudinal electric field to accelerate the ions towards the electron peaks. We develop a simple model that balances the ponderomotive force with the Coulomb force associated with the charge separation. 

A charged particle in an inhomogeneous EM field experiences the ponderomotive force, which is a cycle-averaged force that models the motion of these particles on timescales larger than the laser period. For a particle of mass $m$ with charge $e$ and an electric field with frequency $\omega$ and amplitude {\bf E}, the ponderomotive force is given by
\begin{equation}\label{eq:ponderEq}
    {\bf{F}}_p = -\frac{e^2}{4 m \omega^2} \nabla {\bf E^2(x)},
\end{equation}
where the electric field is cycle-averaged. While this effect is experienced by both electrons and ions, for the laser intensities we are concerned with here, the much more massive ions are hardly affected by the ponderomotive force.

We consider a linearly polarized plane electromagnetic wave propagating in the $+x$ direction and reflecting off of a semi-infinite overdense plasma at $x>0$. Similar to Refs.~\cite{kemp2009hot,may2011mechanism}, we assume that the plasma is a perfect conductor and reflects 100\% of the light, resulting in a standing EM wave (x$<$0) with electric and magnetic fields described by
\begin{eqnarray}\label{eq:sw}
  E_z &= 2 E_0 \sin \left(\frac{2\pi x}{\lambda} \right)\sin(\omega t)  \\
  B_y &= 2 B_0 \cos \left( \frac{2\pi x}{\lambda} \right)\cos(\omega t).
\end{eqnarray}
where $E_0$ is the electric field strength of the laser if there had been no reflection and $B_0 = E_0 / c$.  In real experiments we do not expect 100\% reflectivity. Yet, the reflectivity can be high for laser interactions near, but not significantly above, the threshold for relativistic effects since the high temperatures produce a nearly collisionless plasma, but relativistic absorption is not yet pronounced \cite{levy2014petawatt,Orban_etal2015}. Inserting Eq.~\ref{eq:sw} into Eq.~\ref{eq:ponderEq} yields the longitudinal ponderomotive force associated with the standing wave,
\begin{equation}\label{eq:pf}
     {{F}}_p = -\frac{\lambda e^2 E_0^2}{2 \pi m_e c^2}\sin \left(\frac{4\pi x}{\lambda}\right),
\end{equation}
which will be compared to the Coulomb force associated with the charge separation. 

\subsubsection{Sinusoidal Density Variation Model}
\label{sec:sine}

We begin with a simple model that balances the ponderomotive force on the electrons with the Coulomb force associated with a sinusoidal density variation in the pre-plasma. We assume that before reaching the overdense plasma at $x>0$ the laser travels through a constant, sub-critical density shelf. We estimate the strength and spatial dependence of the electrostatic force in this shelf by choosing a distribution to perfectly balance the ponderomotive force when integrated with the one-dimensional Poisson equation. This produces a sinusoidal electron density modulation of the form
\begin{equation}
    n_{\rm ele} = n_0 + n_e\cos \left(\frac{4\pi x }{ \lambda}\right), \label{eq:density}
\end{equation}
where $n_0$ is the average electron density in the plasma (i.e. the electron density at that location in the plasma before the laser pulse arrives) and $n_e$ describes the amplitude of the density modulation. Equation~\ref{eq:density} is useful for gaining qualitative insight into the ponderomotive steepening process. We remind the reader that the ponderomotive force is time averaged, so this simple model does not fully capture the physics involved. Moreover, as presented in the following sections, simulations indicate that the electron distribution is more strongly peaked than this.

Note that because the local electron density must always be greater than or equal to zero, $n_e$ in Eq.~\ref{eq:density} must not exceed  $n_0$ as you cannot remove more electrons than are available in the plasma. Since the laser only travels in the sub-critical-density region of the plasma, $n_0$ must also be less than the critical density, $n_{\rm crit} = {4\pi^2\varepsilon_o m_e c^2}/{\lambda^2e^2},$ and the maximum electron density is limited. Integrating Eq.~\ref{eq:density} with the one-dimensional\footnote{For experimental beam profiles, we assume that the laser spot size is much larger than $ \lambda/2$. The laser focus is well into the target for our simulations.} Poisson equation results in a quasi-static electric field in the longitudinal (${x}$) direction of the form
\begin{equation}\label{eq:ef}
    E = -\frac{e n_e \lambda}{4\pi \varepsilon_0}\sin \left(\frac{4\pi x}{ \lambda} \right). 
\end{equation}
According to Eq.~\ref{eq:ef} the peak longitudinal electric field is expressed by
\begin{equation}
        E_{\rm max} =  \frac{n_e}{n_{\rm crit}} \frac{\pi m_e c^2}{e \lambda} 
\end{equation}
which is equivalent to
\begin{equation}\label{eq:emax1}
    E_{\rm max} =  (1.6 \times 10^{12} \, {\rm V/m}) \times \left( \frac{n_e}{n_{\rm crit} }\right)  \left( \frac{1 \, \si{\um}}{\lambda}\right). 
 \end{equation}

Note that $n_e$ depends on the intensity of the laser. If we equate the peak ponderomotive force (Eq.~\ref{eq:pf}) to the electrostatic force from (Eq.~\ref{eq:ef}), one finds that for this model the laser is limited to displacing electron densities up to
\begin{align}\label{eq:nmax}
    n_{\rm e, max}&= \frac{4 I }{m_e c^3} \nonumber \\&= (1.6 \times 10^{21} \, {\rm cm^{-3}})\times  \left( \frac{I}{10^{18} \, {\rm W~ cm^{-2}}}  \right) 
\end{align}
where the laser intensity $I= c \varepsilon_0 |E_0|^2/2$ has been used to simplify the expression and $n_{\rm e, max}$ is less than the initial electron density in the plasma. For this model, the density modulation is saturated with a critical intensity of 
\begin{align} \label{eq:critIntensity}
    I_{\rm crit} &= \frac{m_e c^3 n_0}{4} \nonumber  \\&= ( 6.8 \times 10^{17}~ {\rm W ~cm^{-2})} \times \left( \frac{n_0}{n_{\rm crit}}\right)  \left(\frac{1 \si{\um}}{\lambda}\right)^2. 
\end{align}
For this critical intensity, the normalized vector potential $a_0$ for the laser ($a_0 = eE_0 / m_e \omega c$), is 
\begin{equation}
    a_{0,\rm{crit}} = \sqrt{\frac{n_0}{2n_{\rm crit}}},
\end{equation}
or in terms of the electron plasma frequency, $\omega_{pe} = \sqrt{n_e e^2 / m_e \varepsilon_0}$ (using $n_e = n_0$),
\begin{equation}
    a_{0,\rm{crit}} = \sqrt{\frac{1}{2}} \frac{\omega_{pe}}{\omega}.
\end{equation}
Since $a_{0,\rm{crit}} \lesssim 0.7$, it is clear that the applicability of this model does not extend to the strongly relativistic regime. Intensities somewhat above this limit are considered in the next subsection. We note the similarity to this estimate with the wave-breaking limit in laser wake-field acceleration \cite{TajimaDawsonWakefield}, and in \citet{Lehmann2019} this intensity threshold relates to the transition between what they call the ``collective electron" regime to the ``single electron bouncing" regime. For high electron temperatures, this type of model could be extended by considering the Bohm-Gross frequency\cite{BohmGross} like in Ref\cite{Lehmann2019}. Laser driven instabilities would also play a role in certain regimes\cite{kruerbook,drake2010high}.

\subsubsection{Maximum Depletion Limiting Case}
\label{sec:maxdeplete}

At laser intensities significantly above the critical estimate derived in the previous subsection, the electrons are more strongly peaked than predicted by Eq.~\ref{eq:density} and our sinusoidal model breaks down, as demonstrated by simulation results that will be presented later. Although the sinusoidal model breaks down, there is a simple way to determine the maximum longitudinal electric fields in this limiting case. If all of the available electrons are evacuated to the peaks,  the maximum electric field in the maximum depletion regime is a factor of $\pi$ greater than the sinusoidal model. This comes from integrating the charge density in the depletion region ($e n_0 \times \lambda/4)$, giving the maximum longitudinal electric field to be 
\begin{align}        
E_{\rm max} &= \frac{e n_0 }{\varepsilon_0} \left(\frac{\lambda}{4} \right)= \frac{n_0}{n_{\rm crit}} \frac{\pi^2 m_e c^2}{e \lambda} \nonumber \\
&=    (5 \times 10^{12} \, {\rm V/m}) \times \left( \frac{n_e}{n_{\rm crit} }\right)  \left( \frac{1 \, \si{\um}}{\lambda}\right).
\end{align}
This result is notable simply in that it implies that the longitudinal electric field is enhanced (relative to Eq.~\ref{eq:emax1}) at intensities slightly exceeding $I_{\rm crit}$ from Eq.~\ref{eq:critIntensity}, rather than being suppressed.

\subsection{Timescale of the ion acceleration}\label{sec:timescale}
This subsection determines a timescale for ion motion (for ions to reach an electron peak), which will be useful for comparison to the duration of the laser pulse. If the timescale for ion motion is longer than the laser pulse duration, then we characterize this as the `short pulse' regime. If instead, the laser pulse duration is significantly longer than this timescale, we label this as the `long pulse' regime.

We assume the plasma to be an initially neutral mixture of electrons and ions with charge $+Ze$ where $Z$ is the average ionization. Now we consider the electrostatic force on an ion of mass $m_i$ between two of the electron peaks. Following the sinusoidal model, we focus on the ions at a distance of $\lambda/8$ or less away from an electron peak as they will reach the electron peak more quickly (the farthest away ions are considered in Appendix~\ref{ap:maxIonE}), and we approximate the electric field as linear in this region (matching the slope of Eq.~\ref{eq:ef} near its root), or
\begin{equation}\label{eq:force}
    F = -\frac{4Ze^2I}{m_e c^3 \varepsilon_0} x.
\end{equation}
This results in simple harmonic motion with an angular frequency of
\begin{equation}
   \omega_{\rm ion} =  \sqrt{\frac{4 Z e^2 I }{m_i m_e c^3 \varepsilon_0}}. 
\end{equation}
We use this equation to compute the oscillation period of the ion motion. Since we are primarily interested in the dynamics of the ion density peak growth, we are concerned with the timescale for an ion to move from its initial location to the electron density peak. This timescale is equivalent one-quarter of the ion oscillation period (Appendix~\ref{ap:freq}), which is
\begin{equation}\label{eq:timescale}
    \tau_{\rm ion} =\frac{\pi}{4}\sqrt{\frac{m_i m_e \varepsilon_0 c^3}{Ze^2 I}},
\end{equation}
where we note this formula is only valid for $I\le I_{\rm crit}$ where $I_{\rm crit}$ is given by Eq~\ref{eq:critIntensity}. For higher intensities, as discussed in Sec.~\ref{sec:sine}, the maximum electron density that the laser displaces is limited by the number of available electrons and critical density. This results in a minimum timescale of
\begin{align}\label{eq:timescale_min}
    \tau_{\rm ion, min} &=\frac{\pi}{2}\sqrt{\frac{m_i \varepsilon_0}{Ze^2 n_0}}\nonumber \\ &= (51 \, {\rm fs}) \left(\frac{n_{\rm crit}}{n_0} \right)^{1/2} \left( \frac{\lambda}{1~ \si{\um}}\right) \left( \frac{m_{\rm i}}{2 Z m_p} \right)^{1/2}
\end{align}
where in the approximation we have for convenience assumed $Z=1$ and $m_i \approx 2 m_p$ where  $m_p$ is the mass of a proton.

\begin{figure}
\centering
\includegraphics[width=1\linewidth]{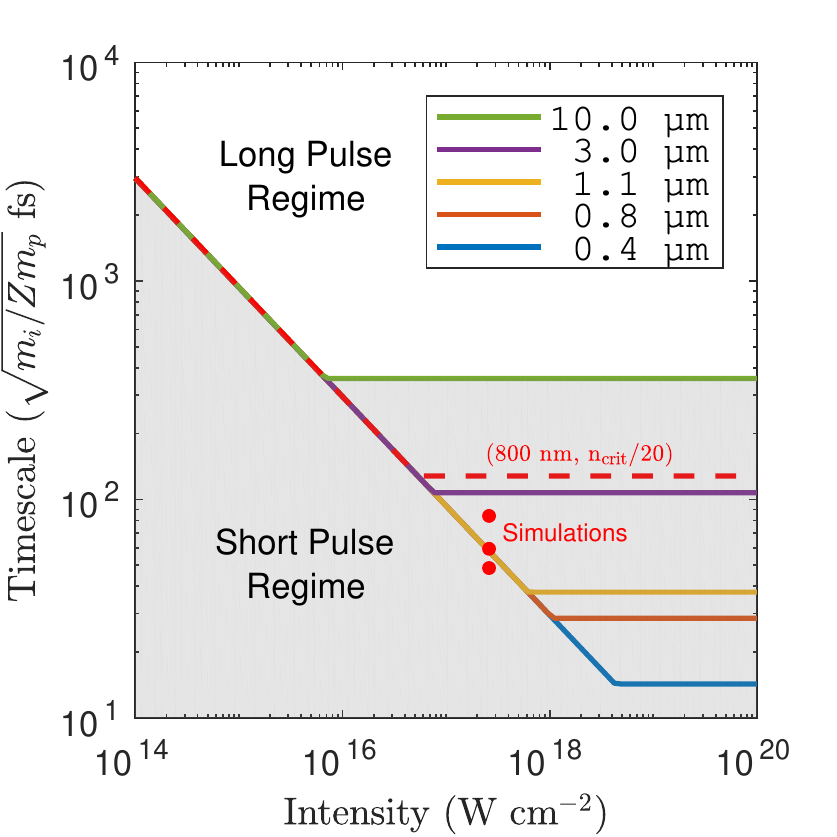}
\vspace{-0.5cm}
\caption{The division between the short pulse and long pulse regime as a function of laser intensity and for a variety of wavelengths (for $n_0 \approx n_{\rm crit}$) for our simple model. The timescale on the vertical axis is the timescale of ion motion from electrostatic forces in ponderomotive steepening. The dashed line represents the timescale for a shelf density of $n_{\rm crit}/20$, pertaining to the simulations in this paper. The individual points on the graph represent the full pulse duration for our simulations (scaled by ion mass).  } \label{fig:estimate}
\end{figure}

Equations~\ref{eq:timescale} and \ref{eq:timescale_min} are represented in Fig.~\ref{fig:estimate} which illustrates the division between the short pulse and long pulse regime as a function of laser intensity and wavelength. The laser wavelength does not appear in Eq.~\ref{eq:timescale}, which is why at low intensities in Fig.~\ref{fig:estimate} the timescale does not depend on wavelength. 
At higher intensities the separation between the two regimes does depend on wavelength because our minimum timescale (Eq.~\ref{eq:timescale_min}) depends on the initial plasma density, where $n_{\rm crit}$ does depend on wavelength. Above this critical intensity, according to the sinusoidal model, the maximum electron density that the laser could displace exceeds the available number of electrons in the plasma near the electron peak. It should be noted that Eq.~\ref{eq:timescale} has the same scaling with parameters as the time estimate in \citet{Lehmann2019} for an ion ``grating" to develop in the standing wave. 

\subsubsection{Maximum ion velocities and energies}\label{sec:maxVelEnergy}
We assumed simple harmonic motion to obtain $\tau_{\rm ion}$ in Eq.~\ref{eq:timescale}. This approach also provides a characteristic ion energy which can be compared to our simulations. Assuming simple harmonic motion with an amplitude of $\lambda/8$ and an available electron density of $n_{\rm e}$ we have an ion velocity that increases with time as
\begin{equation}\label{eq:velion}
    v_{\textnormal{ion}} \approx  \frac{\pi}{4}  \sqrt{\frac{Z m_e}{m_i}}\sqrt{\frac{n_{\rm e}}{n_{\rm crit}}}c ~\sin\left(\frac{\pi}{2}\frac{t_{\rm SW}}{\tau_{ \rm ion}}\right),
\end{equation}
where $t_{\rm SW}$ is time elapsed since the standing wave fields began. Although Eq.~\ref{eq:velion} does not explicitly depend on the laser wavelength, as mentioned earlier our expression for $\tau_{\rm ion}$  is only valid at laser intensities below the critical intensity (Eq.~\ref{eq:critIntensity}) which does depend on wavelength. Since shorter wavelength lasers have a higher critical intensity, one can reach much smaller values of $\tau_{\rm ion}$  as illustrated in Fig.~\ref{fig:estimate}, which would allow the ion velocity (Eq.~\ref{eq:velion}) to grow more quickly. But this growth is limited by the duration of the laser pulse if we are considering the short pulse regime ($t_{\rm SW} \ll \tau_{\rm ion}$).

For a sufficiently long duration laser pulse the standing wave fields will last long enough that $t_{\rm SW}$ approaches $\tau_{\rm ion}$. From Eq.~\ref{eq:velion}, it is straightforward to show that this implies a maximum kinetic energy exceeding 100~keV, 
\begin{align}\label{eq:kemax}
   {\rm{KE}}_{\rm max} &\approx  \frac{\pi^2}{32}Zm_e\frac{n_{\rm e}}{n_{\rm crit}} c^2 \sin^2\left(\frac{\pi}{2}\frac{t_{\rm SW}}{\tau_{ \rm ion}}\right) \nonumber\\ &\approx 157.6\textnormal{~keV} \times Z\left(\frac{n_{\rm e}}{n_{\rm crit}}\right) \sin^2\left(\frac{\pi}{2}\frac{t_{\rm SW}}{\tau_{ \rm ion}}\right) . 
\end{align}
Interestingly, this expression is independent of wavelength except for the wavelength dependence of $n_{\rm crit}$. We note that this model is limited as it neglects motion of ions that are initially further than $\lambda/8$ from an electron peak, which require a longer pulse for maximum energy. We also approximate the field as linear. Alternatively, one could find the maximum energy from the work done by the electric field, this is included in Appendix~\ref{ap:maxIonE}.

\section{Particle-In-Cell Simulations}
\label{sec:sims}
Multiply-peaked ponderomotive steepening is examined numerically with implicit 2D(3$v$) PIC simulations performed with the LSP PIC code\cite{Welch_etal2004}. The initial conditions are such that we are in the short pulse regime of our model and we have exceeded the critical intensity for our model (Fig.~\ref{fig:estimate}). For these simulations, an $x-z$ Cartesian geometry is used, where the laser propagates in the $+x$ direction and the polarization is in the $z$ direction. The simulations have a spatial resolution of 25~nm $\times$ 25~nm ($\lambda/32 \times \lambda/32$)  and were run for 400~fs with a 0.1~fs time step.

\begin{figure}
\includegraphics[width=1\linewidth]{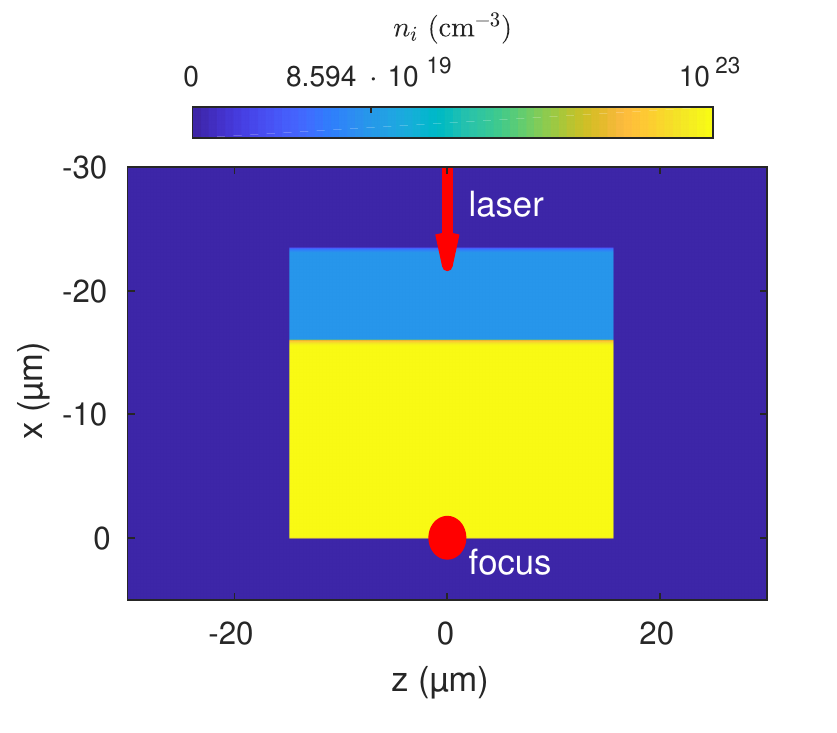}
\caption{Initial conditions for the 2D(3$v$) PIC simulations. The laser propagates in the $+x$ direction with a rectangular target composed of an extended constant under-dense pre-plasma shelf region preceding an overdense region. } \label{fig:init_conditions}
\end{figure}

To isolate the dynamics of the ion peak formation process we consider an idealized geometry of a rectangular target with an extended pre-plasma shelf. The plasma is assumed to be singly ionized with fixed ionization. This choice is made to prevent the critical surface from moving significantly due to ionization caused by the laser pulse. Ponderomotive steepening still occurs in simulations when the critical surface moves forward from ionization (e.g.~Refs.~\cite{Orban_etal2015,ngirmang2017particle}) but we ignore this effect in order to focus on the electron and ion dynamics. In the laser propagation direction, the target consists of a 7~\si{\um} long constant sub-critical density plasma shelf ($n = 8.594 \times 10^{19}$ cm$^{-3}$ $\approx n_{\rm crit}/20$) with a sharp interface between an 15~\si{\um} overdense target ($n = 10^{23}$ cm$^{-3}$ $\approx 60 n_{\rm crit}$) as illustrated in Fig.~\ref{fig:init_conditions}. In the polarization direction, the target is 30~\si{\um} wide. The ions are modeled as collisionless, which is discussed in Sec.~\ref{sec:ion_motion}.

We describe three different simulations with targets composed of fully ionized hydrogen, deuterium, and tritium ions in order to investigate different charge-to-mass ratios. These simulations all keep the laser intensity and initial target electron and ion number densities constant. The overdense region is given a number density similar to  our group's previous work~\cite{Orban_etal2015}. The simulations were initialized with 9 particles per cell for the electrons and 7 particles per cell for the ions with initial thermal energies of 1~eV.

We consider an 800~nm wavelength, normally incident  laser pulse propagating in the $+x$ direction that would reach a peak intensity of \Wcm{18} if no target were present. The pulse duration is 42~fs full width at half maximum (FWHM) with a sine-squared envelope and a Gaussian spot size of 1.5~\si{\um} (FWHM). These parameters are similar to those of the Ti:Sapphire kHz repetition rate laser system described in Refs.~\cite{Morrison_etal2015,Orban_etal2015,Feister_2017}. The laser focus is set at the back of the target, as shown in Fig.~\ref{fig:init_conditions}, in order to create a large region over which ponderomotive steepening can occur. We use these parameters to explore the short pulse regime of our model with $t_{\rm SW} < \tau_{\rm ion}.$


\begin{figure*}
\centering
\includegraphics{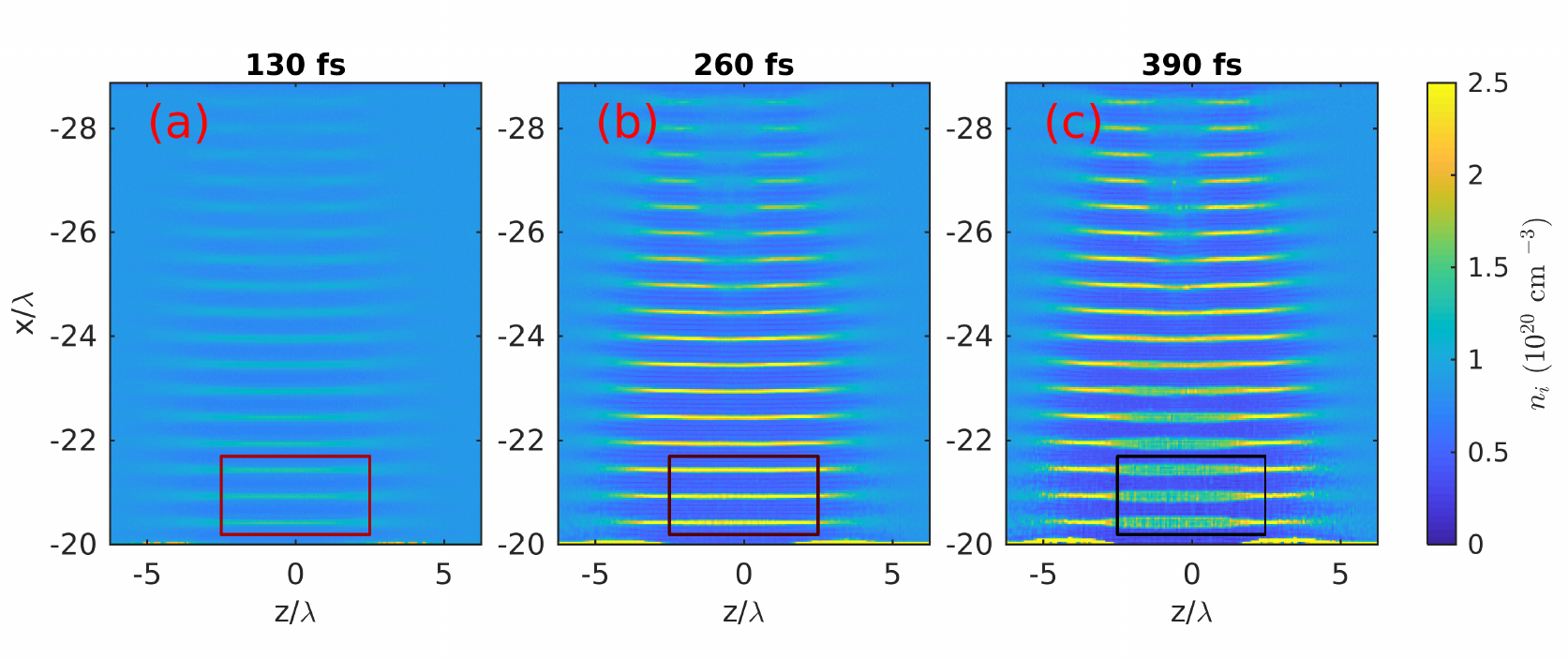}
\vspace{-1.5em}
\caption{Ion density near the reflection point for the deuterium simulation. The laser finishes reflecting around 130~fs from the beginning of the simulation (a), although the peaks continue to grow as shown at 260~fs (b) and then begin to dissipate as illustrated at 390~fs (c). The width of the box (in $z$) represents the region considered in Fig.~\ref{fig:growth_plt} and the entire box represents the region considered for ion trajectories in Fig.~\ref{fig:init_trajectories}. This density peak growth process including the electron density is highlighted in the supplemental video included with this article. } \label{fig:DensitySnapShot}
\end{figure*}
\section{Results}\label{sec:results}

\subsection{Peak Formation and Density Profile Modification}\label{peakFormation}

Figure~\ref{fig:DensitySnapShot} shows snapshots of the ion density from the deuterium simulation at three different times. The standing EM wave causes the electrons to form peaks which, over time, produce peaks in the ion density separated by approximately $\lambda/2$ throughout the under-dense region. The hydrogen and tritium simulations show similar behavior, with the growth of the ion peaks happening sooner for lighter ions and later for more massive ions. In all three simulations we observe more than 10 ion density peaks in the 7~\si{\um} long underdense region. 

\begin{figure}
\includegraphics[width=1\linewidth]{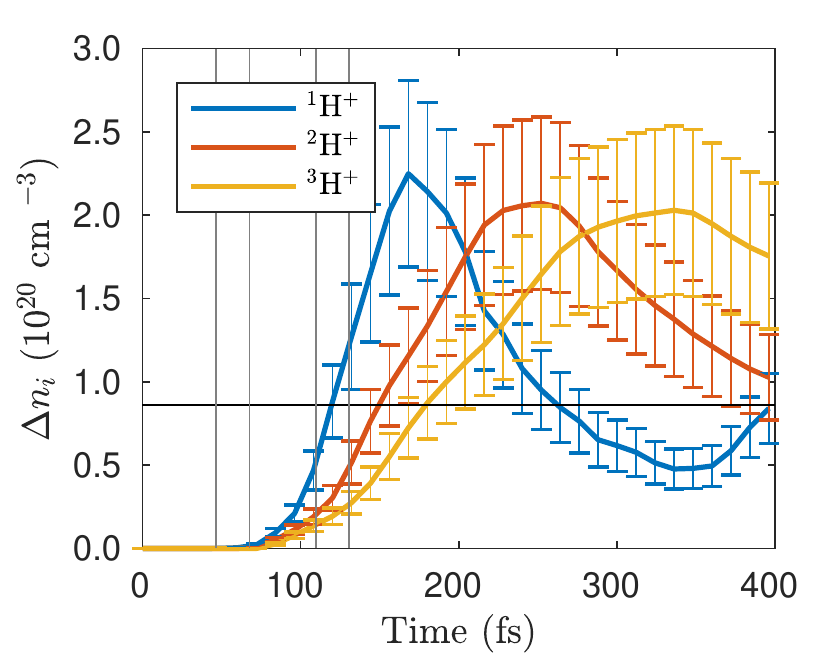}
\vspace{-0.9cm}
\caption{Change in density of the first peak in the ion density. The lines represent the maximum density of the first ion peak averaged over the width of the laser pulse.  This is calculated by averaging the maximum density for each value of $z$ in the region -2 \si{\um} $< z < 2$ \si{\um}, where error bars represent the standard deviation. We note that the exact density at the peak depends on the cell size (and sharpness of the peak), thus this graph comments more on densities in the region near the peak rather than the peak itself. } \label{fig:growth_plt}
\end{figure}

As mentioned, if no target were present, the laser pulse in this simulation would reach \Wcm{18}. Instead, the laser is focused many microns into the target, making the intensity near the sharp interface much lower than it would be in the vacuum case. In our simulations the intensity at the sharp interface is $\approx 2.6 \times$\Wcm{17}. According to Eq.~\ref{eq:critIntensity}, for our wavelength and plasma density $I_{\rm crit} \approx 5 \times$\Wcm{16}. Our simulations therefore explore the regime where the intensity is about five times larger than this threshold.  Regarding the timescale of ion motion, for these simulations $\tau_{\rm ion} = 129$~fs $\times \sqrt{m_{\rm ion}/m_{p}}$. This timescale in all three simulations is longer than the 42~fs FWHM laser pulse (and even the full simulated 84~fs pulse with a sine-squared envelope), making these interactions well within the short pulse regime as illustrated in Fig.~\ref{fig:sketch}.

As discussed in the next section, an examination of the ion trajectories confirm that ions accelerated from both sides of the peak are streaming past each other. Figure~\ref{fig:growth_plt} shows this happening in all three simulations, albeit on different timescales. For each simulation, the peak ion density increases to $\approx 2.5 \times 10^{20}$~cm${}^{-3}$ (approximately three times the initial density), which lasts for tens to hundreds of femtoseconds, and then begins to decrease.  Multiply-peaked ponderomotive steepening in this short pulse regime is therefore a highly transient effect.

\subsection{Ion Motion}\label{sec:ion_motion}
\begin{figure}
\includegraphics[width=1\linewidth]{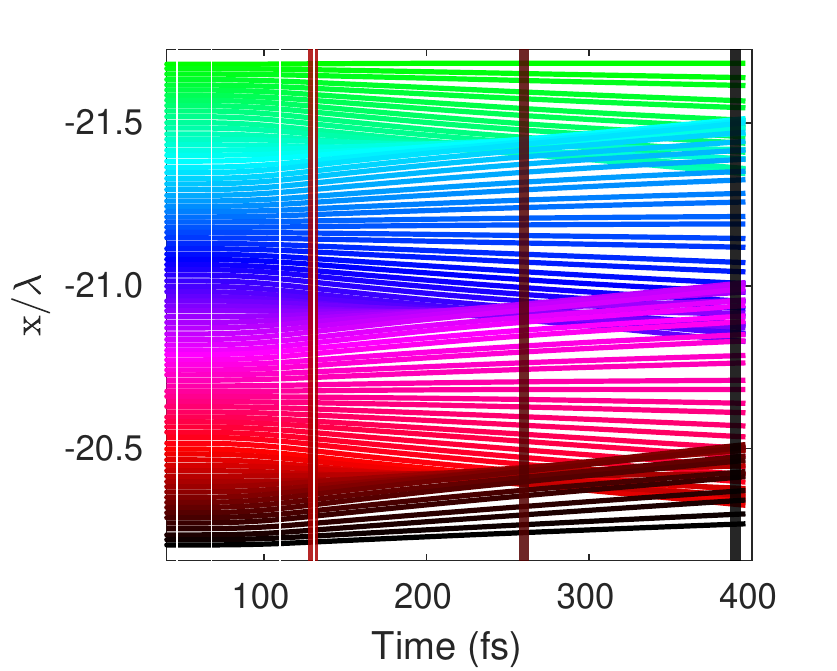}
\caption{The average trajectories in $x$ for a sample of particles starting in the boxed regions in Fig.~\ref{fig:DensitySnapShot}, representing the first three peaks in the ion density. The white vertical lines represent approximately when the laser begins reflecting, reaches its half maxima, and stops reflecting. Shaded vertical lines correspond to the times represented in Fig.~\ref{fig:DensitySnapShot}. The ions continue to travel after the standing wave has dissipated, and the observed peaks are created by the crossing ions.} \label{fig:init_trajectories}
\end{figure}

To better understand the dynamics of the peak formation process, we consider the motion of the ion macroparticles in the simulation. In particular, if we consider the ion trajectories (Fig.~\ref{fig:init_trajectories}) we see that the ions are accelerated towards the electron peaks while the standing wave is present. Later, as the standing wave dissipates, the inertia of the ions allows them to continue to travel with a roughly constant velocity. 
We see from Fig.~\ref{fig:init_trajectories} that many of the ions travel through the peak before the end of the simulation, which produces the broadening observed in Fig.~\ref{fig:DensitySnapShot}. We note that the transverse movement of the ions is 
negligible compared to the longitudinal motion.

\begin{figure}
\includegraphics{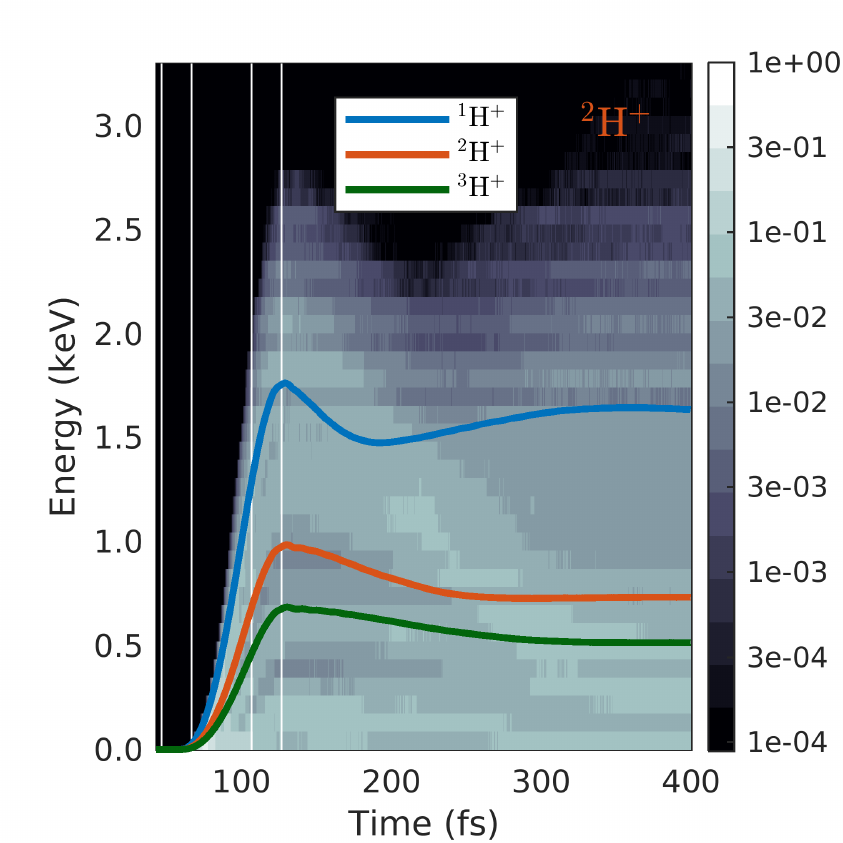}
\caption{Longitudinal ion energies for particles starting in the boxed regions in Fig.~\ref{fig:DensitySnapShot}. The average energies for each simulation are plotted in time and the distribution of ion energies in the background corresponds to the deuterium simulation (logarithmic grayscale). The energies increase while the charge separation caused by standing EM wave is present.  }  \label{fig:ion_energy}
\end{figure}

The energy distribution of the ions is represented in Fig.~\ref{fig:ion_energy} which highlights results from the deuterium simulation and overlays the average ion energies from the hydrogen and tritium simulations. In all three simulations, the ions are accelerated while electron density peaks from the standing wave are present, reaching keV energies. The average ion energy decreases slightly as the standing wave dissipates.  The conversion efficiencies from laser energy to ($>100$~eV) ion energy were approximately $0.027\%,~0.016\%,$ and $0.011\%$ for the three simulations respectively.

We did not include ion-ion collisions in these simulations, which could potentially change the behavior of the ions and potentially lengthen the duration of the peak. However, the peak forms in a relatively low initial density ($8.5 \times 10^{19}$ cm$^{-3}\approx n_{\rm crit} / 20$) plasma shelf. In Appendix~\ref{ap:ionion} we determine that the mean free path of ion-ion collisions for our conditions is larger than the scale of the peak for the higher energy ions in the shelf region.

 \subsection{Peak Electric Fields}
\begin{figure}
\includegraphics[width=1\linewidth]{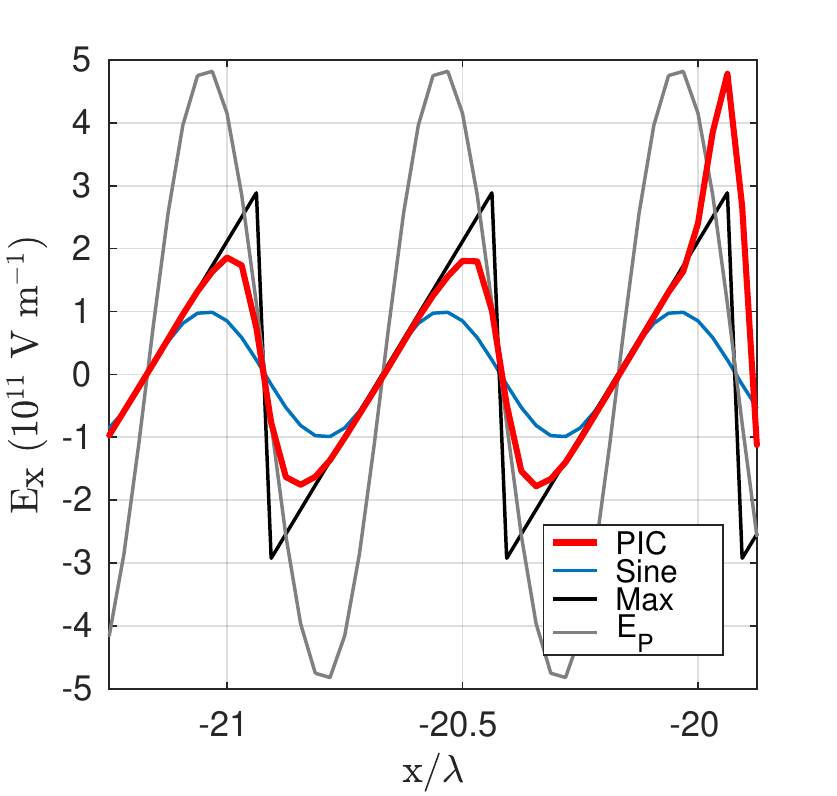}
\caption{The observed longitudinal component of the electric field at 70~fs after the beginning of the deuterium simulation near the center of the laser pulse (PIC) averaged over several cells, as compared to the simple sinusoidal density variation model (Sine), maximum depletion (Max), and the expected ponderomotive force (Eq.~\ref{eq:pf}) divided by $e$ for reference ($E_p$). The electric fields found in the simulation lie between the sinusoidal model and maximum depletion model as expected for this intensity and density. } \label{fig:EFields}
\end{figure}

Figure~\ref{fig:EFields} shows a line out of the longitudinal electric field along the laser axis from the deuterium simulation compared to various models for context. As mentioned, the intensity of this standing wave exceeds $I_{\rm crit}$ by about a factor of 5, which means that we do not expect the sinusoidal model to be accurate in this case. As seen in Fig.~\ref{fig:EFields}, the peak sustained longitudinal electric fields in the simulation are close to $2\times  10^{11}$ V~m$^{-1}$ which is larger than one would expect in this case from the sinusoidal model ($10^{11}$ V~m$^{-1}$) by about a factor of 2. This is still somewhat below the peak electric field of the ``maximum depletion" model shown in Fig.~\ref{fig:estimate} which is near $3.1 \times 10^{11}$~V~m$^{-1}$. This model is described in Sec.~\ref{sec:maxdeplete} and it concludes that the peak electric fields are up to a factor of $\pi$ larger than the sinusoidal model as a limiting case. The results from the simulation lie between these two bounds. At the critical surface, where there are more available electrons, larger fields are present as shown in Fig.~\ref{fig:EFields}, although there are oscillations in the field. When moving further away from the laser axis, there are oscillations in the longitudinal electric field.

\subsection{Ion Energies}

Equation~\ref{eq:kemax} estimates the maximum ion energies from the interaction that we compare to the PIC simulations, however this estimate requires some assumption for how long the standing wave is in place ($t_{\rm SW}$). This is a difficult number to uniquely establish because the intensity envelope of the laser pulse is sine-squared and there is no abrupt turn on and turn off of the standing wave. From considering the results of Fig.~\ref{fig:ion_energy},  using the 42~fs FWHM of the laser pulse as the duration of the standing wave is too short because the ion energies continue to grow even 42~fs after it begins to rise. Using the 84~fs full pulse duration of the laser pulse as the duration of the standing wave is too long, both empirically from Fig.~\ref{fig:ion_energy} and from the reality that the standing waves are created by the overlap of the forward and reflected laser pulse.  In Tab.~\ref{tab:energies} we therefore use both of these timescales for the standing wave in our model in order to bracket the possible ion energies. We empirically find that choosing $t_{\rm SW}$ to be 76~fs yields particularly accurate estimates for the max ion energies in all three simulations.

\begin{table}
\caption{\label{tab:energies}Maximum ion energies reported in keV from the simulation shortly after the standing wave has dissipated. This is compared to the energies predicted with Eq.~\ref{eq:velion}. Because the laser pulse has a temporal profile that is sine squared (rather than square), the time-dependent maximum amplitude makes comparison to the model more ambiguous.  We compare the simulation result to the model with three different assumptions for the duration longitudinal electric field caused by the charge separation from the standing wave ($t_{\rm sw}$).}
\begin{ruledtabular}
\begin{tabular}{lc|c cc}
 &  Simulation &  & Model  &  \\
&  & $t_{\rm sw}=42$~fs &  $t_{\rm sw}=76$~fs&$t_{\rm sw}=84$~fs \\
\hline
$^1\rm H^+$ (keV)&4.9 &1.9& 5.0 & 5.7\\
$^2\rm H^+$ (keV)&3.0 &1.0& 2.9 & 3.4\\
$^3 \rm H^+$ (keV)&2.0 & 0.7& 2.0 & 2.4\\
\end{tabular}
\end{ruledtabular}
\end{table}

\section{Discussion}
 \label{sec:discuss}
 
A multiply-peaked density modulation is observed in our simulations throughout the under-dense shelf region for these initial conditions. The short pulse regime for ponderomotive steepening identified in this theoretical work shows large longitudinal electric fields (potentially up to $\approx~10^{12}$~V~m$^{-1}$ for 800~nm light near the critical density) that accelerate ions to tens to hundreds of keV in energy when the above conditions are satisfied. The consequences of these conditions seem to be overlooked in the literature. From a peak ion energy standpoint, this mechanism is not as appealing as conventional laser-based acceleration schemes such as Target Normal Sheath Acceleration (TNSA)\cite{hatchett2000tnsa}, but because the energies are still sufficient to produce fusion, experiments of this kind may be useful, for example, for producing neutrons with a very small source size.

Largely because the spacing between density peaks is close to $\lambda/2$, features like these have not yet been observed in optical interferometry. By using intense mid-IR laser systems to produce these modulations this may be possible, so long as one is careful to consider that the peaks are a highly transient effect. In our simulations  with 800~nm wavelengths, 42~fs FWHM pulse durations, and peak intensities near $10^{18}$~W~cm$^{-2}$ the features persist for less than a picosecond. While there are interferometric systems that can operate at this short timescale (e.g.~Ref.~\cite{Feister_etal2014}), experiments at longer wavelengths, lower intensities, and ions with lower charge-to-mass ratios can be designed to make the ion acceleration happen over a longer timescale in order to study the evolution of these peaks. The interferometric data would be useful as a novel validation test of kinetic plasma codes, especially if the experiment can be performed at normal incidence.

There are papers in the literature that study the presence of periodicity in the density distribution from the overlap of two crossed laser pulses (e.g.~\citet{suntsov2009femtosecond, Sheng2003}) because this produces a kind of transient ``plasma grating" that can be detected with probe light. The growth of this plasma grating is similar in many ways to the peaks that form via ponderomotive steepening and we outline a number of parallels in the present paper to recent work by \citet{Lehmann2019} who consider overlapping laser pulses through a low density medium. This phenomenon has interesting potential applications as discussed in Refs.~\cite{Lehmann2017, Sheng2003}.

Compared to approaches with counter-propagating laser pulses, there are some advantages to producing these density modulations through the reflection of laser light from an overdense target. Specifically, less total laser energy is required because the reflected laser pulse interferes with itself and there is no need to carefully time the overlap of the pulses since the laser naturally reflects from an overdense surface. The other advantage of overdense targets, as we have explored in this paper, is simply that the density of the shelf or medium the laser travels through can be significantly larger than counter-propagating laser experiments would allow. Larger densities allow for significantly larger longitudinal electric fields for accelerating ions. The density of the medium in experiments with overlapping laser pulses is typically a few orders of magnitude below critical density because of the need to avoid intensity dependent index of refraction effects.  Experiments with overdense targets are not as constrained by this because irradiating an overdense target with an appropriate ``pre-pulse" produces a few-to-many-micron sub-critical density plasma in front of it. Besides increasing the peak ion energies, the other advantage of producing density modulations in a higher density medium is that the difference between the peak and minimum density will be larger, which should produce more easily detectable fringe shifts in efforts to perform interferometric imaging.

We have emphasized the novelty of performing experiments of this kind in the mid-IR ($2$~\si{\um} $\lesssim \lambda \lesssim  10$~\si{\um}). Our results also imply that it would be interesting to investigate ponderomotive steepening with shorter wavelengths as well.  Shorter wavelength lasers are able to propagate into denser regions and, as previously discussed, denser plasmas produce larger peak electric fields which are advantageous for accelerating ions. This detail is important for the possibility of using experiments of this kind to create a neutron source with a very small source size because, as is well known, neutron yields increase significantly with ion energy\cite{davis2008angular}. In a suitably designed experiment, one could try to produce neutrons from the collision of counter-streaming ions in the density peaks. However, as considered in Appendix~\ref{ap:ionion}, the mean free path for these collisions is large compared to $\lambda/2$. Neutron-producing fusion reactions are more likely to come from ions that stream towards the first peak near the target and continue into the overdense region. This would be a ``pitcher-catcher" type configuration where the pitcher and catcher are separated by only $\approx \lambda / 2$.

A crucial assumption of this work is that the plasma remains highly reflective. This is certainly true of our simulations, but it is well known that the intensity and wavelength of the laser are important factors for the reflectivity. To make more reliable extrapolations to shorter and longer wavelengths and smaller and larger intensities than we consider in the simulations we present here, one would need to carefully consider the scaling of the reflectivity with various parameters (e.g.~\citet{levy2014petawatt}). While it is outside the scope of the present work, this remains an important priority for future investigations.

\section{Conclusions}
\label{sec:summary}

The formation of multiply-peaked density modulations associated with ponderomotive steepening is of fundamental interest as a basic plasma process and of practical interest as a means to modify the density profile of a target and to accelerate ions. Our PIC simulations indicate that these peaks are especially transient, lasting less than a picosecond after the end of a short-pulse laser interaction. This is important to factor into the design of future experiments to detect this phenomenon. We also find that the large longitudinal electric fields that are produced in these laser interactions accelerate ions to few keV energies in short pulse laser interactions, and potentially up to hundreds of keV energies in longer duration interactions. In our simulations these fields reach $2\times 10^{11}$~V~m$^{-1}$. 

We outline a simple model to estimate the timescale of ion motion and peak energies of ions in these interactions. This model matches the peak ion energies in our simulations reasonably well. We also comment on extensions to this model that provide some insight even when the laser intensity exceeds a critical value. The model indicates that higher field strengths are achieved with shorter wavelength interactions due to the increased critical density.  Ion acceleration should be much less pronounced in longer wavelength interactions, but this may still be an interesting regime to perform interferometric imaging as a novel validation test of plasma codes if the experiments are performed at normal incidence.

Multiply-peaked ponderomotive steepening has many parallels to studies of counter-propagating laser pulses which is a phenomenon with interesting potential applications for the field\cite{suntsov2009femtosecond,Sheng2003}. A key difference is that interference from reflection occurs at a comparatively higher density. As a result, the longitudinal electric field strengths are much larger, as just mentioned, and there are important subtleties to analytically modeling this phenomenon and challenges in experimentally probing it that we have outlined.

\begin{acknowledgments}

This research is supported by the Air Force Office of Scientific Research under LRIR Project 17RQCOR504 under the management of Dr.~Riq Parra. This project also benefited from a grant of time at the Onyx supercomputer (ERDC) and storage space at the Ohio Supercomputer Center. Support was also provided by the DOD HPCMP Internship Program and the AFOSR summer faculty program.

\end{acknowledgments}
\appendix
\section{Timescale from Ion Oscillation Frequency}\label{ap:freq}
Alternatively, to represent the timescale of the ion motion, we could consider the ion plasma oscillation frequency \begin{equation}
     \omega_{pi} = \sqrt{\frac{{n_i Z^2 e^2}}{{m_i \varepsilon_0}}}.
\end{equation}
We replace $Z n_i$ with $n_{\rm e}$ and then use Eq.~\ref{eq:nmax} to write this as a function of laser intensity,
\begin{equation}
   \omega_{pi} =  \sqrt{\frac{4 Z e^2 I }{m_i m_e c^3 \varepsilon_0}}, 
\end{equation}
and providing a timescale of (one-quarter of the ion oscillation period)
\begin{equation}\label{eq:timescale2}
    \tau_{\rm ion} = \frac{\pi}{4}\sqrt{\frac{m_i m_e \varepsilon_0 c^3}{Ze^2 I}},
\end{equation}
which agrees with the timescale found in Sec.~\ref{sec:timescale}.

\section{Maximum Ion Energies}\label{ap:maxIonE}
To calculate the maximum ion energy, one may also calculate the work done by the electric field on an ion traveling from a valley to an electron peak. For the sinusoidal model, this results in a maximum energy of 
\begin{align}
      \rm{KE}_{\rm max} &= \frac{1}{2} m_e c^2 Z \times \left( \frac{n_e}{n_{\rm crit}}\right) \nonumber \\&= 255.5~Z\times \left( \frac{n_e}{n_{\rm crit}}\right)  \textnormal{keV},
\end{align}
which, as expected, is slightly higher than predicted with the energy predicted from the linear approximation in Eq.~\ref{eq:force}. If $n_e\approx n_0,$ then the field would to shrink at later times, due to ion movement, reducing this maximum energy. 

Similarly, if we calculated the energy from the maximum depletion model, we find  
\begin{align}
\rm{KE}_{\rm max}& = \frac{\pi^2}{8} m_e c^2 Z~\left( \frac{n_e}{n_{\rm crit}}\right) \nonumber \\&= 630.4~Z~\left( \frac{n_e}{n_{\rm crit}}\right) \textnormal{keV},
\end{align}
where we see that the ion acceleration associated with ponderomotive steepening appears to be primarily a sub-MeV acceleration mechanism. 

\section{Ion Mean Free Path}\label{ap:ionion}
Following the ion-ion mean free path estimate from \citet{PARK201238} and \citet{Trubnikov}, the ion-ion mean free path for colliding flows with mass number $A$ and velocity $\bf v$ before the collision is approximated as
\begin{equation}
    \lambda_{\rm mfp} [{\rm cm}] = 5 \times 10^{-13} \left [ {\rm{\frac{s^4}{cm^6}}}\right] \frac{A^2}{Z^4}\frac{{\bf v}^4}{n},
\end{equation}
assuming the Coulomb logarithm is ten\footnote{A more precise treatment could be used in the calculation (e.g. Ref.~\cite{NRL}), but 10 should be sufficient for this approximation.} and that the temperature of the counter streaming flows is much smaller than the energy of the ions due to the bulk flow velocity. For example, if we look at the deuterium simulation, ($A_z= 2, ~Z=1$), and consider the maximum peak density to be $\approx 2.5 \times 10^{20}$~cm$^{-3}$ and average velocity to be $\approx 3\times10^5$~m/s, then we find $\lambda_{\rm mfp}\approx 65$~\si{\um}, which is orders of magnitude larger than the width of the peaks. This mean free path is shorter for the low energy ions, although these are of less interest for this work. For higher densities, such as in the bulk of the target, or with a higher density pre-plasma, collisions would be more significant.

\bibliography{ms}

\begin{thebibliography}{38}
\expandafter\ifx\csname natexlab\endcsname\relax\def\natexlab#1{#1}\fi
\expandafter\ifx\csname bibnamefont\endcsname\relax
  \def\bibnamefont#1{#1}\fi
\expandafter\ifx\csname bibfnamefont\endcsname\relax
  \def\bibfnamefont#1{#1}\fi
\expandafter\ifx\csname citenamefont\endcsname\relax
  \def\citenamefont#1{#1}\fi
\expandafter\ifx\csname url\endcsname\relax
  \def\url#1{\texttt{#1}}\fi
\expandafter\ifx\csname urlprefix\endcsname\relax\def\urlprefix{URL }\fi
\providecommand{\bibinfo}[2]{#2}
\providecommand{\eprint}[2][]{\url{#2}}

\bibitem[{\citenamefont{Estabrook et~al.}(1975)\citenamefont{Estabrook, Valeo,
  and Kruer}}]{estabrook1975two}
\bibinfo{author}{\bibfnamefont{K.}~\bibnamefont{Estabrook}},
  \bibinfo{author}{\bibfnamefont{E.}~\bibnamefont{Valeo}}, \bibnamefont{and}
  \bibinfo{author}{\bibfnamefont{W.}~\bibnamefont{Kruer}},
  \bibinfo{journal}{The Physics of Fluids} \textbf{\bibinfo{volume}{18}},
  \bibinfo{pages}{1151} (\bibinfo{year}{1975}).

\bibitem[{\citenamefont{Kruer}(1988)}]{kruerbook}
\bibinfo{author}{\bibfnamefont{W.~L.} \bibnamefont{Kruer}},
  \emph{\bibinfo{title}{Physics of laser plasma interactions}}
  (\bibinfo{publisher}{Addison-Wesley}, \bibinfo{year}{1988}).

\bibitem[{\citenamefont{Plaja and Roso}(1997)}]{plaja199diffraction}
\bibinfo{author}{\bibfnamefont{L.}~\bibnamefont{Plaja}} \bibnamefont{and}
  \bibinfo{author}{\bibfnamefont{L.}~\bibnamefont{Roso}},
  \bibinfo{journal}{Physical Review E} \textbf{\bibinfo{volume}{56}},
  \bibinfo{pages}{7142} (\bibinfo{year}{1997}).

\bibitem[{\citenamefont{Sheng et~al.}(2003)\citenamefont{Sheng, Zhang, and
  Umstadter}}]{Sheng2003}
\bibinfo{author}{\bibfnamefont{Z.-M.} \bibnamefont{Sheng}},
  \bibinfo{author}{\bibfnamefont{J.}~\bibnamefont{Zhang}}, \bibnamefont{and}
  \bibinfo{author}{\bibfnamefont{D.}~\bibnamefont{Umstadter}},
  \bibinfo{journal}{Applied Physics B} \textbf{\bibinfo{volume}{77}},
  \bibinfo{pages}{673} (\bibinfo{year}{2003}), ISSN \bibinfo{issn}{1432-0649},
  \urlprefix\url{https://doi.org/10.1007/s00340-003-1324-2}.

\bibitem[{\citenamefont{Lehmann and Spatschek}(2016)}]{Lehmann2016}
\bibinfo{author}{\bibfnamefont{G.}~\bibnamefont{Lehmann}} \bibnamefont{and}
  \bibinfo{author}{\bibfnamefont{K.~H.} \bibnamefont{Spatschek}},
  \bibinfo{journal}{Phys. Rev. Lett.} \textbf{\bibinfo{volume}{116}},
  \bibinfo{pages}{225002} (\bibinfo{year}{2016}),
  \urlprefix\url{https://link.aps.org/doi/10.1103/PhysRevLett.116.225002}.

\bibitem[{\citenamefont{Lehmann and Spatschek}(2017)}]{Lehmann2017}
\bibinfo{author}{\bibfnamefont{G.}~\bibnamefont{Lehmann}} \bibnamefont{and}
  \bibinfo{author}{\bibfnamefont{K.}~\bibnamefont{Spatschek}},
  \bibinfo{journal}{Physics of Plasmas} \textbf{\bibinfo{volume}{24}},
  \bibinfo{pages}{056701} (\bibinfo{year}{2017}).

\bibitem[{\citenamefont{Lehmann and Spatschek}(2019)}]{Lehmann2019}
\bibinfo{author}{\bibfnamefont{G.}~\bibnamefont{Lehmann}} \bibnamefont{and}
  \bibinfo{author}{\bibfnamefont{K.}~\bibnamefont{Spatschek}},
  \bibinfo{journal}{Physics of Plasmas} \textbf{\bibinfo{volume}{26}},
  \bibinfo{pages}{013106} (\bibinfo{year}{2019}).

\bibitem[{las()}]{lasermag}
\emph{\bibinfo{title}{Mid-ir lasers: Power and pulse capability ramp up for
  mid-ir lasers}}, \bibinfo{note}{accessed: 2017-1-1},
  \urlprefix\url{http://www.laserfocusworld.com/articles/print/volume-50/issue-05/features/mid-ir-lasers-power-and-pulse-capability-ramp-up-for-mid-ir-lasers.html}.

\bibitem[{\citenamefont{Ngirmang et~al.}(2017)\citenamefont{Ngirmang, Orban,
  Feister, Morrison, Chowdhury, and Roquemore}}]{ngirmang2017particle}
\bibinfo{author}{\bibfnamefont{G.~K.} \bibnamefont{Ngirmang}},
  \bibinfo{author}{\bibfnamefont{C.}~\bibnamefont{Orban}},
  \bibinfo{author}{\bibfnamefont{S.}~\bibnamefont{Feister}},
  \bibinfo{author}{\bibfnamefont{J.~T.} \bibnamefont{Morrison}},
  \bibinfo{author}{\bibfnamefont{E.~A.} \bibnamefont{Chowdhury}},
  \bibnamefont{and}
  \bibinfo{author}{\bibfnamefont{W.}~\bibnamefont{Roquemore}},
  \bibinfo{journal}{Physics of Plasmas} \textbf{\bibinfo{volume}{24}},
  \bibinfo{pages}{103112} (\bibinfo{year}{2017}).

\bibitem[{MUR()}]{MURIresearch}
\emph{\bibinfo{title}{Muri mid-infrared strong-field interaction, research
  thrusts}},
  \bibinfo{howpublished}{\url{http://muri-mir.osu.edu/node/13/##research}},
  \bibinfo{note}{accessed: 2017-1-1}.

\bibitem[{\citenamefont{Fedosejevs et~al.}(1977)\citenamefont{Fedosejevs,
  Tomov, Burnett, Enright, and Richardson}}]{Fedosejevs_etal1977}
\bibinfo{author}{\bibfnamefont{R.}~\bibnamefont{Fedosejevs}},
  \bibinfo{author}{\bibfnamefont{I.~V.} \bibnamefont{Tomov}},
  \bibinfo{author}{\bibfnamefont{N.~H.} \bibnamefont{Burnett}},
  \bibinfo{author}{\bibfnamefont{G.~D.} \bibnamefont{Enright}},
  \bibnamefont{and} \bibinfo{author}{\bibfnamefont{M.~C.}
  \bibnamefont{Richardson}}, \bibinfo{journal}{Phys. Rev. Lett.}
  \textbf{\bibinfo{volume}{39}}, \bibinfo{pages}{932} (\bibinfo{year}{1977}),
  \urlprefix\url{https://link.aps.org/doi/10.1103/PhysRevLett.39.932}.

\bibitem[{\citenamefont{Gong et~al.}(2016)\citenamefont{Gong, Tochitsky, Fiuza,
  Pigeon, and Joshi}}]{Gong_etal2016}
\bibinfo{author}{\bibfnamefont{C.}~\bibnamefont{Gong}},
  \bibinfo{author}{\bibfnamefont{S.~Y.} \bibnamefont{Tochitsky}},
  \bibinfo{author}{\bibfnamefont{F.}~\bibnamefont{Fiuza}},
  \bibinfo{author}{\bibfnamefont{J.~J.} \bibnamefont{Pigeon}},
  \bibnamefont{and} \bibinfo{author}{\bibfnamefont{C.}~\bibnamefont{Joshi}},
  \bibinfo{journal}{Phys. Rev. E} \textbf{\bibinfo{volume}{93}},
  \bibinfo{pages}{061202} (\bibinfo{year}{2016}),
  \urlprefix\url{http://link.aps.org/doi/10.1103/PhysRevE.93.061202}.

\bibitem[{\citenamefont{Suntsov et~al.}(2009)\citenamefont{Suntsov,
  Abdollahpour, Papazoglou, and Tzortzakis}}]{suntsov2009femtosecond}
\bibinfo{author}{\bibfnamefont{S.}~\bibnamefont{Suntsov}},
  \bibinfo{author}{\bibfnamefont{D.}~\bibnamefont{Abdollahpour}},
  \bibinfo{author}{\bibfnamefont{D.}~\bibnamefont{Papazoglou}},
  \bibnamefont{and}
  \bibinfo{author}{\bibfnamefont{S.}~\bibnamefont{Tzortzakis}},
  \bibinfo{journal}{Applied Physics Letters} \textbf{\bibinfo{volume}{94}},
  \bibinfo{pages}{251104} (\bibinfo{year}{2009}).

\bibitem[{\citenamefont{{Morrison} et~al.}(2015)\citenamefont{{Morrison},
  {Chowdhury}, {Frische}, {Feister}, {Ovchinnikov}, {Nees}, {Orban}, {Freeman},
  and {Roquemore}}}]{Morrison_etal2015}
\bibinfo{author}{\bibfnamefont{J.~T.} \bibnamefont{{Morrison}}},
  \bibinfo{author}{\bibfnamefont{E.~A.} \bibnamefont{{Chowdhury}}},
  \bibinfo{author}{\bibfnamefont{K.~D.} \bibnamefont{{Frische}}},
  \bibinfo{author}{\bibfnamefont{S.}~\bibnamefont{{Feister}}},
  \bibinfo{author}{\bibfnamefont{V.~M.} \bibnamefont{{Ovchinnikov}}},
  \bibinfo{author}{\bibfnamefont{J.~A.} \bibnamefont{{Nees}}},
  \bibinfo{author}{\bibfnamefont{C.}~\bibnamefont{{Orban}}},
  \bibinfo{author}{\bibfnamefont{R.~R.} \bibnamefont{{Freeman}}},
  \bibnamefont{and} \bibinfo{author}{\bibfnamefont{W.~M.}
  \bibnamefont{{Roquemore}}}, \bibinfo{journal}{Physics of Plasmas}
  \textbf{\bibinfo{volume}{22}}, \bibinfo{eid}{043101} (\bibinfo{year}{2015}),
  \eprint{1501.02261}.

\bibitem[{\citenamefont{Feister et~al.}(2017)\citenamefont{Feister, Austin,
  Morrison, Frische, Orban, Ngirmang, Handler, Smith, Schillaci, LaVerne
  et~al.}}]{Feister_2017}
\bibinfo{author}{\bibfnamefont{S.}~\bibnamefont{Feister}},
  \bibinfo{author}{\bibfnamefont{D.~R.} \bibnamefont{Austin}},
  \bibinfo{author}{\bibfnamefont{J.~T.} \bibnamefont{Morrison}},
  \bibinfo{author}{\bibfnamefont{K.~D.} \bibnamefont{Frische}},
  \bibinfo{author}{\bibfnamefont{C.}~\bibnamefont{Orban}},
  \bibinfo{author}{\bibfnamefont{G.}~\bibnamefont{Ngirmang}},
  \bibinfo{author}{\bibfnamefont{A.}~\bibnamefont{Handler}},
  \bibinfo{author}{\bibfnamefont{J.~R.~H.} \bibnamefont{Smith}},
  \bibinfo{author}{\bibfnamefont{M.}~\bibnamefont{Schillaci}},
  \bibinfo{author}{\bibfnamefont{J.~A.} \bibnamefont{LaVerne}},
  \bibnamefont{et~al.}, \bibinfo{journal}{Opt. Express}
  \textbf{\bibinfo{volume}{25}}, \bibinfo{pages}{18736} (\bibinfo{year}{2017}),
  \urlprefix\url{http://www.opticsexpress.org/abstract.cfm?URI=oe-25-16-18736}.

\bibitem[{\citenamefont{{Feister} et~al.}(2014)\citenamefont{{Feister}, {Nees},
  {Morrison}, {Frische}, {Orban}, {Chowdhury}, and
  {Roquemore}}}]{Feister_etal2014}
\bibinfo{author}{\bibfnamefont{S.}~\bibnamefont{{Feister}}},
  \bibinfo{author}{\bibfnamefont{J.~A.} \bibnamefont{{Nees}}},
  \bibinfo{author}{\bibfnamefont{J.~T.} \bibnamefont{{Morrison}}},
  \bibinfo{author}{\bibfnamefont{K.~D.} \bibnamefont{{Frische}}},
  \bibinfo{author}{\bibfnamefont{C.}~\bibnamefont{{Orban}}},
  \bibinfo{author}{\bibfnamefont{E.~A.} \bibnamefont{{Chowdhury}}},
  \bibnamefont{and} \bibinfo{author}{\bibfnamefont{W.~M.}
  \bibnamefont{{Roquemore}}}, \bibinfo{journal}{Review of Scientific
  Instruments} \textbf{\bibinfo{volume}{85}}, \bibinfo{eid}{11D602}
  (\bibinfo{year}{2014}), \eprint{1406.3639}.

\bibitem[{\citenamefont{{Grava} et~al.}(2008)\citenamefont{{Grava}, {Purvis},
  {Filevich}, {Marconi}, {Rocca}, {Dunn}, {Moon}, and
  {Shlyaptsev}}}]{Grava_etal2008}
\bibinfo{author}{\bibfnamefont{J.}~\bibnamefont{{Grava}}},
  \bibinfo{author}{\bibfnamefont{M.~A.} \bibnamefont{{Purvis}}},
  \bibinfo{author}{\bibfnamefont{J.}~\bibnamefont{{Filevich}}},
  \bibinfo{author}{\bibfnamefont{M.~C.} \bibnamefont{{Marconi}}},
  \bibinfo{author}{\bibfnamefont{J.~J.} \bibnamefont{{Rocca}}},
  \bibinfo{author}{\bibfnamefont{J.}~\bibnamefont{{Dunn}}},
  \bibinfo{author}{\bibfnamefont{S.~J.} \bibnamefont{{Moon}}},
  \bibnamefont{and} \bibinfo{author}{\bibfnamefont{V.~N.}
  \bibnamefont{{Shlyaptsev}}}, \bibinfo{journal}{Phys.~ Rev.~ E.}
  \textbf{\bibinfo{volume}{78}}, \bibinfo{eid}{016403} (\bibinfo{year}{2008}).

\bibitem[{\citenamefont{{Filevich} et~al.}(2009)\citenamefont{{Filevich},
  {Purvis}, {Grava}, {Ryan}, {Dunn}, {Moon}, {Shlyaptsev}, and
  {Rocca}}}]{Filevich_etal2009}
\bibinfo{author}{\bibfnamefont{J.}~\bibnamefont{{Filevich}}},
  \bibinfo{author}{\bibfnamefont{M.}~\bibnamefont{{Purvis}}},
  \bibinfo{author}{\bibfnamefont{J.}~\bibnamefont{{Grava}}},
  \bibinfo{author}{\bibfnamefont{D.~P.} \bibnamefont{{Ryan}}},
  \bibinfo{author}{\bibfnamefont{J.}~\bibnamefont{{Dunn}}},
  \bibinfo{author}{\bibfnamefont{S.~J.} \bibnamefont{{Moon}}},
  \bibinfo{author}{\bibfnamefont{V.~N.} \bibnamefont{{Shlyaptsev}}},
  \bibnamefont{and} \bibinfo{author}{\bibfnamefont{J.~J.}
  \bibnamefont{{Rocca}}}, \bibinfo{journal}{High Energy Density Physics}
  \textbf{\bibinfo{volume}{5}}, \bibinfo{pages}{276} (\bibinfo{year}{2009}).

\bibitem[{\citenamefont{Morrison et~al.}(2018)\citenamefont{Morrison, Feister,
  Frische, Austin, Ngirmang, Murphy, Orban, Chowdhury, and
  Roquemore}}]{morrison_etal2018}
\bibinfo{author}{\bibfnamefont{J.~T.} \bibnamefont{Morrison}},
  \bibinfo{author}{\bibfnamefont{S.}~\bibnamefont{Feister}},
  \bibinfo{author}{\bibfnamefont{K.~D.} \bibnamefont{Frische}},
  \bibinfo{author}{\bibfnamefont{D.~R.} \bibnamefont{Austin}},
  \bibinfo{author}{\bibfnamefont{G.~K.} \bibnamefont{Ngirmang}},
  \bibinfo{author}{\bibfnamefont{N.~R.} \bibnamefont{Murphy}},
  \bibinfo{author}{\bibfnamefont{C.}~\bibnamefont{Orban}},
  \bibinfo{author}{\bibfnamefont{E.~A.} \bibnamefont{Chowdhury}},
  \bibnamefont{and} \bibinfo{author}{\bibfnamefont{W.~M.}
  \bibnamefont{Roquemore}}, \bibinfo{journal}{New Journal of Physics}
  \textbf{\bibinfo{volume}{20}}, \bibinfo{pages}{022001}
  (\bibinfo{year}{2018}),
  \urlprefix\url{https://doi.org/10.1088%2F1367-2630%2Faaa8d1}.

\bibitem[{\citenamefont{Lee et~al.}(1977)\citenamefont{Lee, Forslund, Kindel,
  and Lindman}}]{Lee_1977}
\bibinfo{author}{\bibfnamefont{K.}~\bibnamefont{Lee}},
  \bibinfo{author}{\bibfnamefont{D.}~\bibnamefont{Forslund}},
  \bibinfo{author}{\bibfnamefont{J.}~\bibnamefont{Kindel}}, \bibnamefont{and}
  \bibinfo{author}{\bibfnamefont{E.}~\bibnamefont{Lindman}},
  \bibinfo{journal}{The Physics of Fluids} \textbf{\bibinfo{volume}{20}},
  \bibinfo{pages}{51} (\bibinfo{year}{1977}).

\bibitem[{\citenamefont{Jones et~al.}(1981)\citenamefont{Jones, Aldrich, and
  Lee}}]{Jones_1981}
\bibinfo{author}{\bibfnamefont{R.~D.} \bibnamefont{Jones}},
  \bibinfo{author}{\bibfnamefont{C.}~\bibnamefont{Aldrich}}, \bibnamefont{and}
  \bibinfo{author}{\bibfnamefont{K.}~\bibnamefont{Lee}}, \bibinfo{journal}{The
  Physics of Fluids} \textbf{\bibinfo{volume}{24}}, \bibinfo{pages}{310}
  (\bibinfo{year}{1981}).

\bibitem[{\citenamefont{Estabrook and Kruer}(1983)}]{Estabrook_1983}
\bibinfo{author}{\bibfnamefont{K.}~\bibnamefont{Estabrook}} \bibnamefont{and}
  \bibinfo{author}{\bibfnamefont{W.}~\bibnamefont{Kruer}},
  \bibinfo{journal}{The Physics of fluids} \textbf{\bibinfo{volume}{26}},
  \bibinfo{pages}{1888} (\bibinfo{year}{1983}).

\bibitem[{\citenamefont{Wenda and Shitong}(1988)}]{wenda1988ponderomotive}
\bibinfo{author}{\bibfnamefont{S.}~\bibnamefont{Wenda}} \bibnamefont{and}
  \bibinfo{author}{\bibfnamefont{Z.}~\bibnamefont{Shitong}},
  \bibinfo{journal}{Physical Review A} \textbf{\bibinfo{volume}{37}},
  \bibinfo{pages}{4387} (\bibinfo{year}{1988}).

\bibitem[{\citenamefont{Kemp et~al.}(2009)\citenamefont{Kemp, Sentoku, and
  Tabak}}]{kemp2009hot}
\bibinfo{author}{\bibfnamefont{A.}~\bibnamefont{Kemp}},
  \bibinfo{author}{\bibfnamefont{Y.}~\bibnamefont{Sentoku}}, \bibnamefont{and}
  \bibinfo{author}{\bibfnamefont{M.}~\bibnamefont{Tabak}},
  \bibinfo{journal}{Physical Review E} \textbf{\bibinfo{volume}{79}},
  \bibinfo{pages}{066406} (\bibinfo{year}{2009}).

\bibitem[{\citenamefont{May et~al.}(2011)\citenamefont{May, Tonge, Fiuza,
  Fonseca, Silva, Ren, and Mori}}]{may2011mechanism}
\bibinfo{author}{\bibfnamefont{J.}~\bibnamefont{May}},
  \bibinfo{author}{\bibfnamefont{J.}~\bibnamefont{Tonge}},
  \bibinfo{author}{\bibfnamefont{F.}~\bibnamefont{Fiuza}},
  \bibinfo{author}{\bibfnamefont{R.}~\bibnamefont{Fonseca}},
  \bibinfo{author}{\bibfnamefont{L.}~\bibnamefont{Silva}},
  \bibinfo{author}{\bibfnamefont{C.}~\bibnamefont{Ren}}, \bibnamefont{and}
  \bibinfo{author}{\bibfnamefont{W.}~\bibnamefont{Mori}},
  \bibinfo{journal}{Physical Review E} \textbf{\bibinfo{volume}{84}},
  \bibinfo{pages}{025401} (\bibinfo{year}{2011}).

\bibitem[{\citenamefont{Levy et~al.}(2014)\citenamefont{Levy, Wilks, Tabak,
  Libby, and Baring}}]{levy2014petawatt}
\bibinfo{author}{\bibfnamefont{M.~C.} \bibnamefont{Levy}},
  \bibinfo{author}{\bibfnamefont{S.~C.} \bibnamefont{Wilks}},
  \bibinfo{author}{\bibfnamefont{M.}~\bibnamefont{Tabak}},
  \bibinfo{author}{\bibfnamefont{S.~B.} \bibnamefont{Libby}}, \bibnamefont{and}
  \bibinfo{author}{\bibfnamefont{M.~G.} \bibnamefont{Baring}},
  \bibinfo{journal}{Nature communications} \textbf{\bibinfo{volume}{5}},
  \bibinfo{pages}{4149} (\bibinfo{year}{2014}).

\bibitem[{\citenamefont{{Orban} et~al.}(2015)\citenamefont{{Orban}, {Morrison},
  {Chowdhury}, {Nees}, {Frische}, and {Roquemore}}}]{Orban_etal2015}
\bibinfo{author}{\bibfnamefont{C.}~\bibnamefont{{Orban}}},
  \bibinfo{author}{\bibfnamefont{J.~T.} \bibnamefont{{Morrison}}},
  \bibinfo{author}{\bibfnamefont{E.~D.} \bibnamefont{{Chowdhury}}},
  \bibinfo{author}{\bibfnamefont{J.~A.} \bibnamefont{{Nees}}},
  \bibinfo{author}{\bibfnamefont{K.}~\bibnamefont{{Frische}}},
  \bibnamefont{and} \bibinfo{author}{\bibfnamefont{W.~M.}
  \bibnamefont{{Roquemore}}}, \bibinfo{journal}{{Physics of Plasmas}}
  (\bibinfo{year}{2015}).

\bibitem[{Note1()}]{Note1}
Note1, \bibinfo{note}{for experimental beam profiles, we assume that the laser
  spot size is much larger than $ \lambda /2$. The laser focus is well into the
  target for our simulations.}

\bibitem[{\citenamefont{Tajima and Dawson}(1979)}]{TajimaDawsonWakefield}
\bibinfo{author}{\bibfnamefont{T.}~\bibnamefont{Tajima}} \bibnamefont{and}
  \bibinfo{author}{\bibfnamefont{J.~M.} \bibnamefont{Dawson}},
  \bibinfo{journal}{Phys. Rev. Lett.} \textbf{\bibinfo{volume}{43}},
  \bibinfo{pages}{267} (\bibinfo{year}{1979}),
  \urlprefix\url{https://link.aps.org/doi/10.1103/PhysRevLett.43.267}.

\bibitem[{\citenamefont{Bohm and Gross}(1949)}]{BohmGross}
\bibinfo{author}{\bibfnamefont{D.}~\bibnamefont{Bohm}} \bibnamefont{and}
  \bibinfo{author}{\bibfnamefont{E.~P.} \bibnamefont{Gross}},
  \bibinfo{journal}{Phys. Rev.} \textbf{\bibinfo{volume}{75}},
  \bibinfo{pages}{1851} (\bibinfo{year}{1949}),
  \urlprefix\url{https://link.aps.org/doi/10.1103/PhysRev.75.1851}.

\bibitem[{\citenamefont{Drake}(2010)}]{drake2010high}
\bibinfo{author}{\bibfnamefont{R.~P.} \bibnamefont{Drake}},
  \bibinfo{journal}{Phys. Today} \textbf{\bibinfo{volume}{63}},
  \bibinfo{pages}{28} (\bibinfo{year}{2010}).

\bibitem[{\citenamefont{{Welch} et~al.}(2004)\citenamefont{{Welch}, {Rose},
  {Clark}, {Genoni}, and {Hughes}}}]{Welch_etal2004}
\bibinfo{author}{\bibfnamefont{D.~R.} \bibnamefont{{Welch}}},
  \bibinfo{author}{\bibfnamefont{D.~V.} \bibnamefont{{Rose}}},
  \bibinfo{author}{\bibfnamefont{R.~E.} \bibnamefont{{Clark}}},
  \bibinfo{author}{\bibfnamefont{T.~C.} \bibnamefont{{Genoni}}},
  \bibnamefont{and} \bibinfo{author}{\bibfnamefont{T.~P.}
  \bibnamefont{{Hughes}}}, \bibinfo{journal}{Computer Physics Communications}
  \textbf{\bibinfo{volume}{164}}, \bibinfo{pages}{183} (\bibinfo{year}{2004}).

\bibitem[{\citenamefont{Hatchett et~al.}(2000)\citenamefont{Hatchett, Brown,
  Cowan, Henry, Johnson, Key, Koch, Langdon, Lasinski, Lee
  et~al.}}]{hatchett2000tnsa}
\bibinfo{author}{\bibfnamefont{S.~P.} \bibnamefont{Hatchett}},
  \bibinfo{author}{\bibfnamefont{C.~G.} \bibnamefont{Brown}},
  \bibinfo{author}{\bibfnamefont{T.~E.} \bibnamefont{Cowan}},
  \bibinfo{author}{\bibfnamefont{E.~A.} \bibnamefont{Henry}},
  \bibinfo{author}{\bibfnamefont{J.~S.} \bibnamefont{Johnson}},
  \bibinfo{author}{\bibfnamefont{M.~H.} \bibnamefont{Key}},
  \bibinfo{author}{\bibfnamefont{J.~A.} \bibnamefont{Koch}},
  \bibinfo{author}{\bibfnamefont{A.~B.} \bibnamefont{Langdon}},
  \bibinfo{author}{\bibfnamefont{B.~F.} \bibnamefont{Lasinski}},
  \bibinfo{author}{\bibfnamefont{R.~W.} \bibnamefont{Lee}},
  \bibnamefont{et~al.}, \bibinfo{journal}{Physics of Plasmas}
  \textbf{\bibinfo{volume}{7}}, \bibinfo{pages}{2076} (\bibinfo{year}{2000}).

\bibitem[{\citenamefont{Davis and Petrov}(2008)}]{davis2008angular}
\bibinfo{author}{\bibfnamefont{J.}~\bibnamefont{Davis}} \bibnamefont{and}
  \bibinfo{author}{\bibfnamefont{G.}~\bibnamefont{Petrov}},
  \bibinfo{journal}{Plasma Physics and Controlled Fusion}
  \textbf{\bibinfo{volume}{50}}, \bibinfo{pages}{065016}
  (\bibinfo{year}{2008}).

\bibitem[{\citenamefont{Park et~al.}(2012)\citenamefont{Park, Ryutov, Ross,
  Kugland, Glenzer, Plechaty, Pollaine, Remington, Spitkovsky, Gargate
  et~al.}}]{PARK201238}
\bibinfo{author}{\bibfnamefont{H.-S.} \bibnamefont{Park}},
  \bibinfo{author}{\bibfnamefont{D.}~\bibnamefont{Ryutov}},
  \bibinfo{author}{\bibfnamefont{J.}~\bibnamefont{Ross}},
  \bibinfo{author}{\bibfnamefont{N.}~\bibnamefont{Kugland}},
  \bibinfo{author}{\bibfnamefont{S.}~\bibnamefont{Glenzer}},
  \bibinfo{author}{\bibfnamefont{C.}~\bibnamefont{Plechaty}},
  \bibinfo{author}{\bibfnamefont{S.}~\bibnamefont{Pollaine}},
  \bibinfo{author}{\bibfnamefont{B.}~\bibnamefont{Remington}},
  \bibinfo{author}{\bibfnamefont{A.}~\bibnamefont{Spitkovsky}},
  \bibinfo{author}{\bibfnamefont{L.}~\bibnamefont{Gargate}},
  \bibnamefont{et~al.}, \bibinfo{journal}{High Energy Density Physics}
  \textbf{\bibinfo{volume}{8}}, \bibinfo{pages}{38 } (\bibinfo{year}{2012}),
  ISSN \bibinfo{issn}{1574-1818},
  \urlprefix\url{http://www.sciencedirect.com/science/article/pii/S1574181811000978}.

\bibitem[{\citenamefont{{Trubnikov}}(1965)}]{Trubnikov}
\bibinfo{author}{\bibfnamefont{B.~A.} \bibnamefont{{Trubnikov}}},
  \bibinfo{journal}{Reviews of Plasma Physics} \textbf{\bibinfo{volume}{1}},
  \bibinfo{pages}{105} (\bibinfo{year}{1965}).

\bibitem[{Note2()}]{Note2}
Note2, \bibinfo{note}{a more precise treatment could be used in the calculation
  (e.g. Ref.~\cite {NRL}), but 10 should be sufficient for this approximation.}

\bibitem[{\citenamefont{{Huba}}(2013)}]{NRL}
\bibinfo{author}{\bibfnamefont{J.~D.} \bibnamefont{{Huba}}},
  \bibinfo{journal}{Naval Research Laboratory}  (\bibinfo{year}{2013}).

\end{thebibliography}
\bibliographystyle{apsrev}

\end{document}